\documentclass[a4paper,11pt]{article}
\usepackage{pos}

\newcommand{\oo}{\mathring}
\newcommand{\abs}[1]{\lvert #1 \rvert}

\usepackage{subfigure}
\usepackage{array}
\usepackage{multirow}
\usepackage{booktabs} 

\title{Noncommutative Quasinormal Modes and the Violation of Isospectrality}
\author[a]{Nikola~Herceg}
\author[a]{Tajron~Jurić}
\author*[a]{A.~Naveena~Kumara}
\author[a]{Andjelo~Samsarov}
\author[b]{Ivica~Smolić}

\affiliation[a]{Rudjer Bo\v{s}kovi\'c Institute,\\ Bijeni\v{c}ka cesta 54, HR-10002 Zagreb, Croatia}
\affiliation[b]{Department of Physics, Faculty of Science, University of Zagreb,\\ Bijeni\v{c}ka cesta 32, 10000 Zagreb, Croatia}

\emailAdd{nherceg@irb.hr}
\emailAdd{tjuric@irb.hr}
\emailAdd{nathith@irb.hr}
\emailAdd{asamsarov@irb.hr}
\emailAdd{ismolic@phy.hr}

\abstract{
We explore quasinormal modes (QNMs) of the Schwarzschild black hole under a noncommutative (NC) deformation of spacetime, constructed via a Drinfeld twist formalism. In this approach, the usual Regge--Wheeler (axial) and Zerilli (polar) equations acquire additional contributions that depend on the NC parameter. Employing semi-analytical approximations (high-order WKB,  Pöschl--Teller and Rosen--Morse), we calculate the corresponding QNM spectra. Our results show that whereas the commutative case preserves the isospectrality of axial and polar modes, noncommutativity systematically violates this degeneracy. The discrepancy grows with the strength of the NC parameter, becoming evident through distinct real and imaginary parts in the ringdown frequencies. These findings highlight the potential of black hole QNMs to serve as probes of quantum-spacetime corrections in strong-field regimes.
}

\FullConference{
}

\begin{document}
\maketitle

\section{Introduction}
Recent breakthroughs in observational astrophysics have firmly established black holes as accessible probes of the strong-gravity regime. On the one hand, the direct detection of gravitational waves (GWs) from coalescing black hole binaries by LIGO/Virgo collaborations has offered clear evidence of black hole mergers and ringdowns, confirming decades-old theoretical predictions of General Relativity (GR)~\cite{Abbott:2016blz, TheLIGOScientific:2016src, Abbott:2016nmj}. On the other hand, horizon-scale imaging, most famously by the Event Horizon Telescope (EHT), has provided the first silhouette (“shadow”) of a supermassive black hole~\cite{Akiyama:2019bqs, Akiyama:2019cqa,  EventHorizonTelescope:2022wkp}. These developments not only place black hole studies on a rigorous observational footing but also highlight the need for ever more precise theoretical models to interpret future data. Despite these successes, fundamental puzzles remain, particularly concerning the role of quantum effects in very strong curvature regions or near the horizon.

One of the most powerful theoretical tools to probe black hole dynamics is the study of \textit{quasinormal modes} (QNMs)~\cite{Kokkotas:1999bd,Berti:2009kk,Konoplya:2011qq}. These modes characterize how a black hole responds to small disturbances, and are effectively the “ringing” frequencies and damping timescales of the geometry. Formally, they arise from solutions to the linearized Einstein equations about a black hole background, with boundary conditions that allow radiation to escape at infinity and to be purely ingoing at the horizon. Because these QNMs depend only on the black hole’s mass, spin, and (if present) electric charge, they effectively serve as “fingerprints” for black hole spacetimes. In astrophysical scenarios, QNMs govern the late-time behavior of emitted GWs, and their precise measurements could shed light on fundamental questions—ranging from the stability of black hole solutions to the validity of exotic horizon-scale effects.

While classical GR successfully predicts both black hole shadows and ringdown signals, a consistent quantum theory of gravity remains elusive, prompting a host of theoretical scenarios for extending GR. One intriguing avenue is \emph{noncommutative gravity}, wherein spacetime geometry itself is modified at very short length scales. Within this approach, fields and coordinate functions cease to commute, typically signifying that the usual notion of a smooth manifold breaks down near the Planck scale. A mathematically elegant way to incorporate such noncommutative (NC) deformations is via the Drinfeld twist formalism, which preserves a generalized notion of diffeomorphism invariance while altering the product structure of tensor fields~\cite{Aschieri:2005yw, Aschieri:2005zs, Aschieri:2009qh}.

In this contribution, we investigate how such a NC deformation affects the QNM spectrum of the Schwarzschild black hole. Specifically, we adopt a newly proposed NC Einstein manifold condition to derive deformed analogs of the Regge-Wheeler (axial) and Zerilli (polar) master equations~\cite{Herceg:2023zlk, Herceg:2023pmc, Herceg:2024vwc}. Classically, these two equations yield the same quasinormal frequencies—an isospectrality property that is a hallmark of Schwarzschild perturbations in GR~\footnote{For early studies on Schwarzschild black hole perturbations, refer to Refs. \cite{Regge:1957td, Edelstein:1970sk, Zerilli:1970se}. Kip Thorne’s chapter, “Introduction to Regge and Wheeler: ‘Stability of a Schwarzschild Singularity’” in \cite{Castellani:2019pvh}, offers both a historical perspective and contemporary insights into black hole perturbation theory and its connection with quasinormal modes.}. Our results show that the NC correction splits the spectrum, breaking this degeneracy between axial and polar modes. We compute these NC QNMs via several semi-analytical methods, including a high-order WKB approximation, and compare the accuracy and convergence of the results. We conclude that, NC gravity provides a compelling phenomenological testbed for quantum-gravitational effects in the strong-field regime—an approach that both extends classical perturbation theory and challenges us to look for subtle deviations from the predictions of GR.

\section{Black Hole Perturbation Theory in Noncommutative Geometry}
% \subsection{Noncommutative Differential Geometry}
In this section, we briefly summarize the main ideas behind black hole perturbations in a NC gravitational framework, as developed in~\cite{Herceg:2023lmt, Herceg:2023pmc, Herceg:2023zlk}. The theoretical foundation for NC gravity employed here is built on the Hopf algebra formalism \cite{Aschieri:2005yw, Aschieri:2005zs, Aschieri:2009qh}. In particular, the Moyal-type deformations considered here arise from the twist
\begin{equation}\label{twist}
	\mathcal{F}=\text{exp}\left[- i \Theta^{\alpha\beta}\partial_{\alpha}\otimes\partial_{\beta}\right]=f^{A}\otimes f_{A},
\end{equation}
with inverse \(\mathcal{F}^{-1}=\bar{f}^A\otimes \bar{f}_A\), where summation over $A$ is implied. The matrix \(\Theta^{\alpha\beta}\) is a constant and antisymmetric, while \(\{\partial_\mu\}\) is the standard coordinate basis for vector fields. This twist deforms the usual product of coordinate functions into the NC algebra
\begin{equation}
[x^{\mu}\stackrel{\star}{,}x^{\nu}]=x^{\mu}\star x^{\nu}-x^{\nu}\star x^{\mu}=i \Theta^{\mu\nu},
\end{equation}
where the \(\star\)-product is 
\begin{equation}
f\star g=fg+i\Theta^{\alpha\beta}\frac{\partial f}{\partial x^{\alpha}}\frac{\partial g}{\partial x^{\beta}}+\mathcal{O}(\Theta^2), \quad \forall \ f,g\in\mathcal{C}^{\infty}.
\end{equation}

The cornerstone of the NC perturbation framework is the notion of a NC Einstein manifold, introduced via \(\mathcal{R}\)-symmetrized Ricci tensor \cite{Herceg:2023lmt}:
\begin{equation}  
\label{NCRsymm}
	\hat{{\rm R}}_{\mu\nu}\equiv\frac{1}{2}\left\langle dx^{\alpha}, \hat{R} (\partial_{\alpha}, \partial_{\mu}, \partial_{\nu})+\hat{R} (\partial_{\alpha}, \bar{R}^{A}(\partial_\nu), \bar{R}_{A}(\partial_\mu) )\right\rangle_\star.
\end{equation}
Here, $\hat{R}$ denotes the $\star$-curvature tensor, 
\begin{equation}
       \hat R (\partial _\mu , \partial _\nu , \partial _\rho)=  \ \hat R_{~\mu \nu \rho}^\sigma \partial_\sigma
	    =\left(\partial _\mu \hat \Gamma _{~\nu \rho}^{ \sigma} -\partial _\nu \hat \Gamma _{~\mu \rho}^{ \sigma} + \hat \Gamma _{~\nu \rho}^{ \tau} \star \hat  \Gamma _{~\mu \tau}^{ \sigma}-\hat \Gamma _{~\mu \rho}^{ \tau} \star \hat \Gamma _{~\nu \tau}^{ \sigma} \right) \partial _\sigma.
\end{equation}
In practical calculations, one starts with a standard (commutative) metric \(g_{\mu \nu}\) and then calculates the NC inverse metric which allows us to obtain the NC Levi-Civita connection (which is both torsion-free and metric-compatible\footnote{This is not the same condition as for a classical metric. Covariant derivative of the metric vanishes, but that of the metric inverse does not.}). They are given by:
\begin{equation}
g^{\star\alpha\beta}=g^{\alpha\beta}-g^{\gamma\beta}i \Theta^{ \mu \nu }(\partial_ \mu g^{\alpha\sigma})(\partial_ \nu g_{\sigma\gamma})+\mathcal{O}(\Theta^2).
\end{equation}
\begin{equation}\label{sLC}
\hat \Gamma_{\mu\nu}^{ \rho} =\frac{1}{2}g^{\star\rho\sigma}\star\left(\partial_{\mu}g_{\nu\sigma}+\partial_{\nu}g_{\mu\sigma}-\partial_{\sigma}g_{\mu\nu}\right).
\end{equation}

For the Moyal–Weyl twist, \(\mathcal{R}\)-symmetrization reduces to ordinary symmetrization in appropriate basis:
\begin{equation}\label{NCR}
	\hat{{\rm R}}_{\mu\nu}=\hat{R} _{(\mu\nu)} \equiv \frac{1}{2} (\hat{R} _{\mu\nu} + \hat{R} _{\nu\mu}).
\end{equation}
% We consider an Abelian twist, a form of the Moyal-Weyl twist where globally defined vector fields $\{ X_\mu \}$ generating the twist commute. This twist has the form
% \begin{equation} \label{abelian}
% 	\mathcal{F}=\exp \left( -i \Theta ^{\mu \nu} X _\mu \otimes X _\nu  \right).
% \end{equation}

For our study, we consider linearized perturbations of a static black hole metric. Specifically, let \( g_{\mu\nu} \;=\;\oo{g}_{\mu\nu} \;+\;h_{\mu\nu}, \) where \(\oo{g}_{\mu\nu}\) is the background and \(h_{\mu\nu}\) is a small perturbation. In particular, we focus on the Schwarzschild geometry,
\begin{equation}
   ds^2 = \oo g_{\mu \nu} dx^{\mu} dx^{\nu} = -\left(1- \frac{R}{r}\right) dt^2 + \left(1- \frac{R}{r}\right)^{-1} dr^2 + r^2\left(d\theta^2 + \sin^2 \theta d\varphi^2\right), 
\end{equation}
where $R = 2M$ denotes the horizon radius, and $M$ is the mass of the  black hole.  We adopt what is known as a semi-pseudo-Killing twist, which is formed by combining the Killing field \(K\) associated with the background Schwarzschild metric \(\oo g_{\mu\nu}\) and an arbitrary
vector field \(X\). Concretely, we define
\begin{equation} \label{semipseudo}
    \mathcal{F} = \text{exp}\left[ -i \frac{a}{2} \big(K \otimes X - X \otimes K \big) \right].
\end{equation}
For a static, spherically symmetric Schwarzschild spacetime, the relevant Killing fields are \(\partial_t\) and \(\partial_\varphi\). We set
\begin{equation} \label{kxdef}
    K = \alpha \partial_t + \beta \partial_\varphi, \qquad X = \partial_r, \qquad \alpha, \beta \in \mathbb{R}.
\end{equation}
As a consequence, the coordinates obey the NC relations
\begin{align*}
	[t\stackrel{\star}{,} r] = i a \alpha, \qquad
	[\varphi \stackrel{\star}{,}r] = i a \beta.
\end{align*}
Next, we introduce a parameter \(\lambda\) by noting that the Killing vector \(K\) acts on the black-hole perturbation \(h_{\mu\nu}\) through the Lie derivative,
\begin{align}
	h_{\mu \nu} \propto e^{i m \varphi}e^{- i \omega t} \implies \pounds_K h_{\mu \nu} = i \lambda \: h_{\mu \nu}, 
	\text{ where } \quad K = \alpha \partial_t + \beta \partial_\varphi, \,  \text{so that} \quad
	\lambda = -\alpha \omega + \beta m. \label{alfabeta}
\end{align}
Since our treatment is linearized in both the metric perturbation and the NC parameter \(a\), we employ the following linearized \(\star\)-product:
\begin{equation}
		h \star k =  h \ k + \  \frac{i}{2} \ a \ h \left[\left(\alpha \overleftarrow{\partial}_t + \beta \overleftarrow{\partial}_\varphi \right) \overrightarrow{\partial}_r - \overleftarrow{\partial}_r \left(\alpha \overrightarrow{\partial}_t + \beta \overrightarrow \partial_\varphi \right) \right] \ k + \mathcal{O}(a^2). 
\end{equation}

\subsection{Axial perturbations and noncommutative Regge-Wheeler potential} \label{sec3A}
We now give an overview of how axial (odd-parity) perturbations are treated in the NC framework outlined above, following the results of \cite{Herceg:2023lmt, Herceg:2023pmc, Herceg:2023zlk}. Working in the Regge–Wheeler gauge, the axial perturbations are parameterized by specific radial functions \cite{Regge:1957td, Langlois:2021xzq}. Concretely, these can be written as
\begin{eqnarray} \label{ansatz}
	&& h_{t \theta}=\frac{1}{\sin \theta} \sum_{\ell, m} h_0^{\ell m} \partial_{\phi} Y_{\ell m}(\theta, \phi)e^{-i \omega t},\quad
	 h_{t \phi}=-\sin \theta \sum_{\ell, m} h_0^{\ell m} \partial_\theta Y_{\ell m}(\theta, \phi)e^{-i \omega t},\\
	&& h_{r \theta}=\frac{1}{\sin \theta} \sum_{\ell, m} h_1^{\ell m} \partial_{\phi} Y_{\ell m}(\theta, \phi) e^{-i \omega t},\quad
	 h_{r \phi}=-\sin \theta \sum_{\ell, m} h_1^{\ell m} \partial_\theta Y_{\ell m}(\theta, \phi) e^{-i \omega t}.
\end{eqnarray}
Substituting these into the NC vacuum equation \(\hat{{\rm R}}_{\mu\nu}=0\) produces a total of 7 non-vanishing equations. These equations are separable into radial and angular parts, leaving 3 distinct radial equations of which only 2 are independent. Hence, by selecting any two of the three radial equations, one obtains the solution of the entire system. After certain algebraic manipulations (see \cite{Herceg:2023zlk} for details), there emerges a single second-order differential equation specifically governing \(h_1\):
\begin{equation}\begin{split}
	&r (r-R) \Big( \ell(\ell + 1) r (R-r) + 2r^2 - 6rR + 5R^2 + \omega^2 r^4 \Big) h_1 +r^2(r - R)^2\Big( (5R - 2r)h_1' + r(r - R)h_1''\Big) \\
 &+ \lambda a \bigg[\Big(\ell(\ell + 1)r(r - R)^2 - 6r^3 + \frac{R}{2}(49
r^2 - 64 r R + 26 R^2 - \omega^2 r^4) \Big) h_1 \\
& + r(r - R)^2 \Big( 3 ( r
- 2R) h_1' + \frac{1}{2}r R h_1''\Big) \bigg]=0. 
\end{split}\end{equation}
After some algebraic steps, one may recast this radial equation into a Schrödinger-like form by introducing both a field redefinition and a deformed tortoise coordinate. Specifically, one defines
\begin{equation}
h_1(r) = \frac{r^2}{r - R}\Big[ 1 + \frac{\lambda a}{2} \Big( \frac{3}{r} - \frac{1}{r - R} + \frac{1}{R} \log \frac{r}{r - R} \Big) \Big]\psi(r)
\end{equation}
together with the NC analog of the tortoise coordinate,
\begin{equation} \label{RWtortoise}
\hat{r}_* = r + R \log \frac{r - R}{R} + \frac{\lambda a}{2} \frac{R}{r - R}.
\end{equation}
The resulting differential equation takes the familiar form
\begin{equation}\label{Rschrodinger}
	\frac{d^2 \psi}{d{\hat r}_*^2} + \Big( \omega^2 - V(r) \Big) \psi = 0,
\end{equation}
where the effective potential can be split into a standard Regge–Wheeler part and an NC correction, 
\begin{equation}\begin{split}\label{RWpotential}
V(r) = V_\text{RW}(r) + V_\text{NC}(r) = \frac{(r - R)\big(\ell (\ell + 1)r - 3R\big)}{r^4} + \lambda a \frac{\ell(\ell + 1)(3R - 2r)r + R(5r - 8R)}{2 r^5}.
\end{split}
\end{equation}
In the commutative limit \(a \to 0\), the correction \(V_\text{NC}\) vanishes, leaving only the usual Regge–Wheeler potential of \cite{Regge:1957td, Edelstein:1970sk}. One may observe, moreover, that setting \(\alpha=0\) and \(\beta=1\) in \eqref{alfabeta} causes \(V_\text{NC}\) to depend on both \(\ell\) and \(m\), leading to a Zeeman-like splitting in the QNM spectrum.

\subsection{Polar perturbations and noncommutative Zerilli potential} \label{sec3B}
We turn next to the polar (or even-parity) sector. In the Zerilli gauge, these perturbations can be expressed in terms of four distinct radial functions \(H_{0}^{\ell m}\), \(H_{1}^{\ell m}\), \(H_{2}^{\ell m}\), and \(K^{\ell m}\). Concretely, one writes \cite{Herceg:2024vwc}
\begin{eqnarray}
\label{eq:even-pert2}
&&h_{tt} = f(r)\sum_{\ell, m} H_{0}^{\ell m}(r) Y_{\ell m}(\theta,\varphi)e^{-i \omega t},  \quad
h_{tr} = \sum_{\ell, m} H_{1}^{\ell m}(r) Y_{\ell m}(\theta,\varphi)e^{-i \omega t}, \\
&&h_{rr} = \frac{1}{f(r)} \sum_{\ell, m} H_{2}^{\ell m}(r) Y_{\ell m}(\theta,\varphi)e^{-i \omega t},  \quad 
h_{ab} = \sum_{\ell, m} K^{\ell m}(r) \oo g_{ab} Y_{\ell m}(\theta,\varphi)e^{-i \omega t},
\end{eqnarray}
where $f(r)=1-R/r$. Indices \(a\) and \(b\) here denote the angular directions \(\theta\) and \(\varphi\). By inserting these expressions into the NC vacuum equation derived from \eqref{NCR}, one obtains 10 radial equations in total, though only 7 of them are distinguishable. 

After performing suitable algebraic manipulations (see \cite{Herceg:2024vwc} for details), it is found that the problem can be encoded in the following pair of coupled first-order equations (where one sets \(L = H_1 / \omega\)):
\begin{equation}
    \begin{aligned} \label{coupled_ode}
        K' &= \left[\alpha _0 (r) + \alpha _2 (r) \omega ^2\right] K + \left[\beta _0 (r) + \beta _2 (r) \omega ^2\right] L, \\
        L' &= \left[\gamma _0 (r) + \gamma _2 (r) \omega ^2\right] K + \left[\delta _0 (r) + \delta _2 (r) \omega ^2\right] L,
    \end{aligned}
\end{equation}
where \(\alpha,\beta,\gamma,\delta\) are somewhat lengthy functions of \(r\), and we refer to \cite{Herceg:2024vwc} for their full forms. To transform these coupled equations into a Schrödinger-type equation, we apply a field redefinition,
\begin{equation} \label{transformation}
        K = \ \hat{f}(r)\hat{K}+\hat{g}(r)\hat{L}, \qquad
        L = \ \hat{h}(r)\hat{K}+\hat{l}(r)\hat{L},
\end{equation}
together with a suitable coordinate transformation \(d r / d \hat r_* = \hat n(r)\). Our goal is to choose \(\hat{f}, \hat{g}, \hat{h}, \hat{l},\) and \(\hat{n}\) such that the following conditions are satisfied:
\begin{equation} \label{requirement}
    \begin{aligned}
	    \frac{d\hat{K}}{d\hat r_*}=\hat{L}, \qquad \frac{d\hat{L}}{d\hat r_*}=(V-\omega ^2)\hat{K},
    \end{aligned}
\end{equation}
which then combine to yield a Schrödinger-like wave equation for \(\hat{K}\):
\begin{equation} \label{Zschrodinger}
    \begin{aligned}
       \frac{d^2\hat{K}}{d\hat r^2_*}+(\omega ^2 -V)\hat{K}=0.
    \end{aligned}
\end{equation}
The final step is to solve for \(\hat{f}, \hat{g}, \hat{h},\) \(\hat{l},\) and \(\hat{n}\) such that the system meets the conditions outlined above. In practice, the functions emerge in the form
\begin{equation} \label{eqfghl}
    \begin{aligned}
	    \hat{f}(r)=&\ f(r)- \, \lambda a \, \tilde f(r), \qquad \hat{g}(r)= g(r)- \, \lambda a \, \tilde g(r),\\
	    \hat{h}(r)=&\ h(r)- \, \lambda a \, \tilde h(r), \qquad \hat{l}(r)= l(r)- \, \lambda a \, \tilde l(r),
    \end{aligned}
\end{equation}
where \(\{f(r),\,g(r),\,h(r),\,l(r)\}\) constitute the commutative contributions, and \(\{\tilde f(r),\,\tilde g(r),\,\tilde h(r),\,\tilde l(r)\}\) are the leading NC corrections. One also defines the NC extension of the tortoise coordinate,
\begin{equation} \label{Ztortoise}
	\hat r_* =\ r + R \log \frac{r - R}{R} - \lambda a \left[\frac{(2 \Lambda +7) R}{2 (2 \Lambda +3)
   (r-R)}-\frac{4 (\Lambda +3) \log
   \left(\frac{r}{R}-1\right)}{(2 \Lambda +3)^2}-\frac{9 \log
   \left(\frac{2 \Lambda  r}{R}+3\right)}{\Lambda  (2 \Lambda
	+3)^2}\right]
\end{equation}
where we use \(\ell(\ell+1)=2\,\Lambda+2\). The resulting Schr{\"o}dinger-like form of the equation \eqref{Zschrodinger} features an effective potential expressed as $V=V_\text{Z}+\, V_\text{NC}$. The term \(V_Z\) coincides with the usual Zerilli potential in the commutative limit, while \(V_\text{NC}\) encodes NC corrections. Concretely,
\begin{equation} \label{Zpotential}
    \begin{aligned}
        V_\text{Z}=&\ \frac{(r-R) \left(8 \Lambda ^2 (\Lambda +1) r^3+12 \Lambda ^2 r^2 R+18 \Lambda  r R^2+9 R^3\right)}{r^4 (2 \Lambda  r+3 R)^2},\\
        V_\text{NC}=&\ \frac{\lambda a}{4 r^5 (2 \Lambda  r+3 R)^3} \Big[32 \Lambda ^2 \left(2 \Lambda ^2+7\right) r^5-8 \Lambda ^2 (2 \Lambda  (6 \Lambda -13)+121) r^4 R  \\
        & \quad -12 \Lambda  (2 \Lambda  (15 \Lambda -58)+59) r^3 R^2-2 (\Lambda  (440 \Lambda -741)+162) r^2 R^3 \\ & \quad -3 (316 \Lambda -207) r R^4-387 R^5 \Big].
    \end{aligned}
\end{equation}

\section{Noncommutative Quasinormal Modes}
\subsection{WKB Method}
We now turn to computing QNM frequencies in the NC framework for both axial and polar perturbations, making use of semi-analytical methods. Originally introduced by Schutz and Will for black hole scattering \cite{Schutz:1985km} and subsequently developed to higher orders \cite{Iyer:1986np, Konoplya:2003ii, Matyjasek:2017psv}, the WKB technique relies on matching solutions at asymptotic infinity and near the horizon by Taylor-expanding around the potential’s maximum between the two turning points. Reformulating the relevant Schrödinger-like equation in the form
\begin{equation}\label{WKB_wave}
\frac{d^2\Psi}{dx^2} + Q(x)\Psi(x) = 0,
\end{equation}
where \(x=r_*\) and \(Q(x)=\omega^2-V\), one finds that if \(-\,Q_{\max}\ll Q(\pm\infty)\), it suffices to approximate the solution near the peak of the potential via
\begin{equation} \label{WKB_Taylor}
Q(x) = Q_0 + \frac{1}{2}Q_0^{\prime \prime}(x - x_0)^2 + \mathcal{O}((x - x_0)^3),
\end{equation}
with \(x_0\) the location of the maximum, \(Q_0=Q(x_0)\), and \(Q_0''\) its second derivative. Substituting \eqref{WKB_Taylor} into \eqref{WKB_wave} results in a parabolic cylinder differential (Weber) equation. The higher-order WKB solution thus generates QNM frequencies through the general form
\begin{equation}
    \frac{i(\omega ^2-V_0 )}{\sqrt{-2V_0^{\prime \prime}}}-\sum _{i=2}^{N} \tilde \Lambda _i=n+\frac{1}{2},
\end{equation}
where each \(\tilde{\Lambda}_i\) depends on derivatives of the potential \(V_0\) and $n$ is the overtone number. As the WKB expansion is asymptotic, increasing the order does not necessarily guarantee greater accuracy; the optimal order is determined by the specific shape of the potential. Its numerical error can be estimated by comparing solutions at consecutive orders \cite{Konoplya:2019hlu},
\begin{equation}
    \Delta _k=\frac{| \omega _{k+1}-\omega _{k-1}|}{2},
\end{equation}
often referring to the fundamental QNM (\(n=0\)) for consistency.

For the NC potentials in \eqref{RWpotential} and \eqref{Zpotential}, we must specify the type of twist chosen; we will fix \(\alpha=0\) and \(\beta=1\) in \eqref{kxdef}, as this simplifies the analysis. Under this choice, only the radial and angular coordinates fail to commute, specifically \([\varphi\stackrel{\star}{,}r]=ia\). Hence, wherever \(\lambda a\) appears in the previous formulae, it can be replaced by \(am\), following \eqref{alfabeta}.

\begin{figure}[t]
\centering
\includegraphics[scale=0.75]{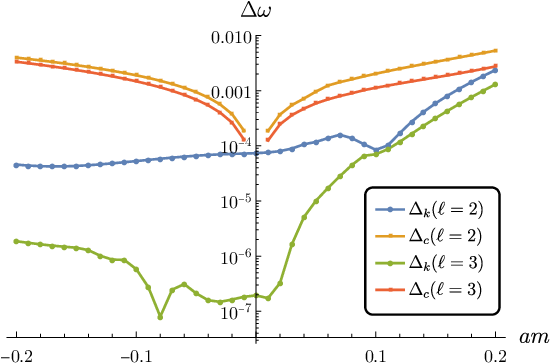}
\includegraphics[scale=0.75]{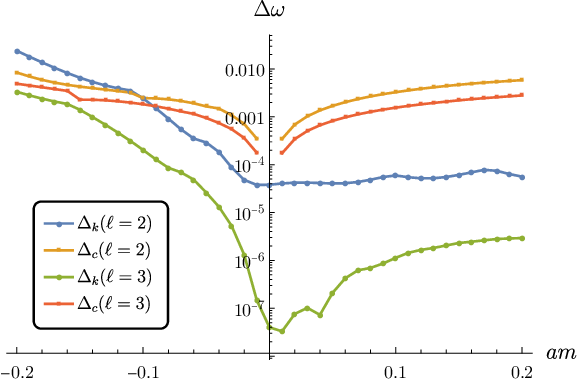}
\caption{Comparison of the NC correction \(\Delta_c\) with the WKB error \(\Delta_k\) at the optimal WKB order. \emph{Left:} Axial modes. \emph{Right:} Polar modes.}
\label{figerror}
\end{figure}

For the standard, commutative versions of both axial and polar potentials, performing a sixth-order WKB approximation generally yields the best agreement for the \(\ell=2\) mode. In the NC setting, however, the optimal WKB order can differ depending on the value of \(a m\). Tables \ref{tab1}--\ref{tab2} list the QNM frequencies at these optimal orders. To assess how the intrinsic WKB error compares against the NC shifts, we define
\begin{equation}
    \Delta _{c} = | \omega _{NC} -\omega_C |,
\end{equation}
where \(\omega\) is the fundamental (lowest-overtone) QNM frequency. We compute \(\Delta_c\) for a range of \(a\,m\) and compare it to the WKB relative error \(\Delta_k\). Figure \ref{figerror} illustrates this comparison for both axial and polar modes.

For higher multipoles, such as \(\ell=3\) and above, the numerical error in the WKB approach is much smaller than for \(\ell=2\). As \(\abs{a m}\) increases, the WKB error \(\Delta_k\) eventually becomes comparable to the NC shift \(\Delta_c\). Nonetheless, within the region where WKB remains reliable, the NC corrections exceed the intrinsic approximation error, making them physically significant. For larger positive values of \(a m\) where the WKB expansion begins to fail, additional semi-analytic methods  can be employed to compute the NC corrections.

%%%%%%%%%%%%%%%%%%%%%%%%%%%%%%%%%%%%%%
%%%%%%%%%%%%%%%%%%%%%%%%%%%%%%%%%%%%%%
\begin{table}[ht!]
\begin{tabular}{m{3em} m{10em} m {3em} m{9em} m{8em}}
 $a m  $  & WKB  & Order & P\"oschl-Teller & Rosen-Morse  \\
\hline
\multicolumn{5}{c}{\textbf{\( \ell=2 \)}} \\
\hline
 $-0.2 $ & 0.3775(61) - 0.0883(97) i & 6 & 0.382049 - 0.090320 i & 0.38335 - 0.08924 i \\
 $-0.1 $ & 0.3755(14) - 0.0887(70) i & 6 & 0.380114 - 0.090466 i & 0.38057 - 0.09008 i \\
 $-0.01$ & 0.3738(07) - 0.0888(92) i & 6 & 0.378454 - 0.090521 i & 0.37855 - 0.09044 i \\
 $-0.001$& 0.3736(38) - 0.0888(91) i & 6 & 0.378294 - 0.090520 i & 0.37838 - 0.09044 i \\
 $ 0   $ & 0.3736(19) - 0.0888(91) i & 6 & 0.378276 - 0.090520 i & 0.37837 - 0.09044 i \\
 $0.001$ & 0.3736(01) - 0.0888(91) i & 6 & 0.378258 - 0.090520 i & 0.37838 - 0.09042 i \\
 $ 0.01$ & 0.3734(33) - 0.0888(88) i & 6 & 0.378099 - 0.090518 i & 0.37836 - 0.09030 i \\
 $ 0.1 $ & 0.3715(87) - 0.0889(38) i & 4 & 0.376562 - 0.090455 i & 0.37756 - 0.08961 i \\
 $ 0.2 $ & 0.36(8345) - 0.08(8195) i & 4 & 0.375007 - 0.090238 i & 0.37611 - 0.08930 i \\
\hline
\multicolumn{5}{c}{\textbf{\( \ell=3 \)}} \\
\hline
 $-0.2 $  & 0.602768 - 0.092381 i & 8  & 0.605558 - 0.093206 i & 0.60880 - 0.09161 i \\
 $-0.1 $  & 0.600920 - 0.092632 i & 10 & 0.603847 - 0.093344 i & 0.60489 - 0.09283 i \\
 $-0.01$  & 0.599573 - 0.092706 i & 8  & 0.602548 - 0.093361 i & 0.60271 - 0.09328 i \\
 $-0.001$ & 0.599456 - 0.092703 i & 8  & 0.602432 - 0.093358 i & 0.60257 - 0.09329 i \\
 $ 0   $  & 0.599443 - 0.092703 i & 8  & 0.602420 - 0.093358 i & 0.60257 - 0.09328 i \\
 $0.001$  & 0.599431 - 0.092702 i & 8  & 0.602407 - 0.093358 i & 0.60259 - 0.09327 i \\
 $ 0.01$  & 0.599318 - 0.092696 i & 12 & 0.602295 - 0.093355 i & 0.60272 - 0.09315 i \\
 $ 0.1 $  & 0.5983(44) - 0.0924(52) i & 4  & 0.601353 - 0.093234 i & 0.60302 - 0.09241 i \\
 $ 0.2 $  & 0.59(6979) - 0.09(1480) i & 4  & 0.600761 - 0.092919 i & 0.60246 - 0.09208 i \\
\hline
\multicolumn{5}{c}{\textbf{\( \ell=4 \)}} \\
\hline
 $-0.2 $ & 0.812255 - 0.093889 i & 11 & 0.814338 - 0.094360 i & 0.81927 - 0.09259 i \\
 $-0.1 $ & 0.810458 - 0.094097 i & 12 & 0.812664 - 0.094503 i & 0.81421 - 0.09394 i \\
 $-0.01$ & 0.809279 - 0.094166 i & 12 & 0.811533 - 0.094537 i & 0.81175 - 0.09446 i \\
 $-0.001$& 0.809188 - 0.094164 i & 12 & 0.811442 - 0.094535 i & 0.81163 - 0.09447 i \\
 $ 0   $ & 0.809178 - 0.094164 i & 12 & 0.811433 - 0.094535 i & 0.81163 - 0.09446 i \\
 $0.001$ & 0.809169 - 0.094164 i & 12 & 0.811423 - 0.094535 i & 0.81167 - 0.09445 i \\
 $ 0.01$ & 0.809084 - 0.094159 i & 12 & 0.811338 - 0.094533 i & 0.81191 - 0.09433 i \\
 $ 0.1 $ & 0.8085(45) - 0.0939(66) i & 4  & 0.810770 - 0.094427 i & 0.81302 - 0.09361 i \\
 $ 0.2 $ & 0.808(397) - 0.093(169) i & 4  & 0.810860 - 0.094128 i & 0.81311 - 0.09332 i \\
\hline
\end{tabular}
\caption{NC axial QNM frequencies for \(n=0\), \(M=1\) (\(R=2\)), and \(\ell=2,3,4\). The values are obtained through high-order WKB, Pöschl–Teller, and Rosen–Morse approximations. “Order” indicates the optimal WKB order. Parenthetical values denote estimated WKB errors.}
\label{tab1}
\end{table}
%%%%%%%%%%%%%%%%%%%%%%%%%%%%%%%%%%%%%%
%%%%%%%%%%%%%%%%%%%%%%%%%%%%%%%%%%%%%%
\begin{table}[ht!]
\begin{tabular}{m{3em} m{10em} m {3em} m{9em} m{8em}}
 $a m  $  & WKB  & Order & P\"oschl-Teller & Rosen-Morse  \\
\hline
\multicolumn{5}{c}{\textbf{\( \ell=2 \)}} \\
\hline
$-0.2$  & 0.3(80198) - 0.0(83646) i & 3 & 0.382642 - 0.097531 i & 0.38379 - 0.09648 i \\
 $-0.1$  & 0.37(4735) - 0.09(1148) i & 4 & 0.380292 - 0.093609 i & 0.38178 - 0.09230 i \\
 $-0.01$ & 0.3738(64) - 0.0892(07) i & 5 & 0.378475 - 0.090866 i & 0.37890 - 0.09050 i \\
 $-0.001$& 0.3736(58) - 0.0889(67) i & 5 & 0.378308 - 0.090622 i & 0.37845 - 0.09050 i \\
 $0$     & 0.3736(36) - 0.0889(40) i & 5 & 0.378290 - 0.090595 i & 0.37839 - 0.09051 i \\
 $0.001$ & 0.3736(13) - 0.0889(14) i & 5 & 0.378272 - 0.090567 i & 0.37836 - 0.09049 i \\
 $0.01$  & 0.3734(13) - 0.0886(75) i & 5 & 0.378109 - 0.090322 i & 0.37821 - 0.09023 i \\
 $0.1$   & 0.3718(88) - 0.0861(75) i & 6 & 0.376612 - 0.088102 i & 0.37741 - 0.08744 i \\
 $0.2$   & 0.3711(29) - 0.0836(58) i & 7 & 0.375215 - 0.085959 i & 0.37794 - 0.08379 i \\
\hline
\multicolumn{5}{c}{\textbf{\( \ell=3 \)}} \\
\hline
 $-0.2$  & 0.60(4367) - 0.09(2871) i & 3  & 0.605859 - 0.096241 i & 0.60775 - 0.09528 i \\
 $-0.1$  & 0.600(941) - 0.093(756) i & 4  & 0.603790 - 0.094691 i & 0.60594 - 0.09361 i \\
 $-0.01$ & 0.599571 - 0.092828 i & 9  & 0.602530 - 0.093487 i & 0.60308 - 0.09321 i \\
 $-0.001$& 0.599456 - 0.092716 i & 10 & 0.602430 - 0.093377 i & 0.60263 - 0.09328 i \\
 $0$     & 0.599443 - 0.092703 i & 10 & 0.602419 - 0.093365 i & 0.60257 - 0.09329 i \\
 $0.001$ & 0.599431 - 0.092690 i & 10 & 0.602408 - 0.093353 i & 0.60255 - 0.09328 i \\
 $0.01$  & 0.599324 - 0.092577 i & 10 & 0.602314 - 0.093235 i & 0.60248 - 0.09315 i \\
 $0.1$   & 0.598589 - 0.091402 i & 9  & 0.601588 - 0.092150 i & 0.60286 - 0.09153 i \\
 $0.2$   & 0.598402 - 0.090106 i & 8  & 0.601184 - 0.091006 i & 0.60541 - 0.08897 i \\
\hline
\multicolumn{5}{c}{\textbf{\( \ell=4 \)}} \\
\hline
 $-0.2$  & 0.813(422) - 0.094(494) i & 3  &  0.814702 - 0.096084 i & 0.81716 - 0.09519 i \\
 $-0.1$  & 0.8104(89) - 0.0947(78) i & 4  &  0.812601 - 0.095290 i & 0.81527 - 0.09432 i \\
 $-0.01$ & 0.809269 - 0.094238 i & 10 &  0.811514 - 0.094611 i & 0.81218 - 0.09437 i \\
 $-0.001$& 0.809187 - 0.094171 i & 12 &  0.811440 - 0.094548 i & 0.81170 - 0.09445 i \\
 $0$     & 0.809178 - 0.094164 i & 12 &  0.811432 - 0.094541 i & 0.81163 - 0.09447 i \\
 $0.001$ & 0.809170 - 0.094156 i & 12 &  0.811424 - 0.094534 i & 0.81161 - 0.09447 i \\
 $0.01$  & 0.809097 - 0.094088 i & 12 &  0.811357 - 0.094463 i & 0.81158 - 0.09438 i \\
 $0.1$   & 0.808733 - 0.093367 i & 11 &  0.810973 - 0.093790 i & 0.81270 - 0.09317 i \\
 $0.2$   & 0.808973 - 0.092531 i & 10 &  0.811068 - 0.093024 i & 0.81676 - 0.09100 i \\
\end{tabular}
\caption{NC polar QNM frequencies for \(n=0\), \(M=1\) (\(R=2\)), and \(\ell=2,3,4\). The values are obtained through high-order WKB, Pöschl–Teller, and Rosen–Morse approximations. “Order” indicates the optimal WKB order. Parenthetical values denote estimated WKB errors.}
\label{tab2}
\end{table}

%%%%%%%%%%%%%%%%%%%%%%%%%%%%%%%%%%%%%%%%%%%%%%
%%%%%%%%%%%%%%%%%%%%%%%%%%%%%%%%%%%%%%%%%%%%%
\subsection{P{\"o}schl-Teller and Rosen-Morse method}
In this subsection, we evaluate NC QNM frequencies by employing the P{\"o}schl–Teller technique and its refinement via the Rosen–Morse potential. The P{\"o}schl–Teller approach approximates the effective potential in our Schr{\"o}dinger-like radial equation with a well-known functional form~\cite{Mashhoon:1982im, Ferrari:1984ozr, Ferrari:1984zz, Blome:1981azp}. Consequently, one replaces the exact potential with
\begin{equation} \label{PT_diff}
	\frac{\partial^2 \Psi}{\partial r_*^2}+\left[ \omega ^2 -\frac{V_0}{\cosh ^2\alpha (r_*-  \bar r_*)} \right] \Psi =0,
\end{equation}
where \(\alpha^2=-\,V_0''/(2\,V_0)\) and \(\bar{r}_*\) denotes the position of the potential peak in the tortoise coordinate, with \(V_0=V(\bar{r}_*)\). Converting this to hyper-geometric form and examining the asymptotic behavior of the solution yields a closed-form expression for the QNM frequencies \cite{Berti:2009kk},
\begin{equation}\label{PTomega}
    \omega = \pm \sqrt{V_0-\frac{\alpha ^2}{4}}-i \alpha \left( n+\frac{1}{2} \right).
\end{equation}

The P{\"o}schl–Teller method can be improved by switching to a more flexible model known as the Rosen–Morse potential \cite{Heidari:2023yjd, Heidari:2023egu}, which includes an additional term to introduce asymmetry and thus provide a more accurate fit to the black hole effective potential. Concretely, one replaces the P{\"o}schl–Teller form with~\cite{PhysRev.42.210}
\begin{equation}
	V_{RM}=   \frac{V_0}{\cosh ^2\alpha (r_* -  \bar r_*)} + V_1 \tanh \alpha (r_*-  \bar r_* ),
\end{equation}
where \(V_1\) encapsulates the asymmetry. Inserting this into the radial wave equation and enforcing suitable boundary conditions produces the QNM frequencies
\begin{equation}
 \frac{\sqrt{\omega ^2 +V_1}+\sqrt{\omega ^2 -V_1} }{ 2}  = \pm \sqrt{V_0-\frac{\alpha ^2}{4}}-i \alpha \left( n+\frac{1}{2} \right).
\end{equation}
In the limit \(V_1 \to 0\), one reverts to the P{\"o}schl–Teller formula \eqref{PTomega}. The QNM frequencies obtained through these P{\"o}schl–Teller and Rosen–Morse approaches are shown in Tables \ref{tab1} and \ref{tab2}, alongside the results from the WKB analysis. This provides an independent cross-check of the NC corrections assessed via the WKB method.

\section{Violation of Classical Isospectrality}
A notable feature of QNMs in classical general relativity is their \emph{isospectrality}: although axial and polar perturbations of a Schwarzschild black hole obey different master equations, they yield identical QNM spectra in the commutative case. This property also holds for Reissner–Nordström, Kerr, and (to linear order) Kerr–Newman black holes \cite{Pani:2013ija, Pani:2013wsa}. In contrast, the NC Schwarzschild scenario exhibits a clear breakdown of isospectrality when we examine the numerical QNM frequencies.

Table~\ref{tab0a} illustrates this for \(a m=0\), where the commutative axial and polar sectors indeed share the same frequencies:
% \begin{table}[h!]
% \begin{tabular}{m{4em} m{10em} m{10em} m {10em} }
% %\hline
% \\
%  & $\ell=2$ & $\ell=3$ &  $\ell=4$ \\
% %\\
% \hline
% \\
% Axial & 0.3736(19) - 0.0888(91) i & 0.599443 - 0.092703 i & 0.809178 - 0.094164 i \\
% Polar & 0.3736(36) - 0.0889(40) i & 0.599443 - 0.092703 i & 0.809178 - 0.094164 i \\
% \\
% \hline
% \end{tabular}
% \caption{QNM frequencies (WKB method) in the commutative limit \(a m=0\), for \(n=0\) and \(M=1\,(R=2)\).}
% \label{tab0a}
% \end{table}
% %
% \par
\begin{table}[h!]
\begin{tabular}{m{4em} m{10em} m{10em} m{10em}}
 & $\ell=2$ & $\ell=3$ & $\ell=4$ \\
\hline
Axial & 0.3736(19) - 0.0888(91)\,i & 0.599443 - 0.092703\,i & 0.809178 - 0.094164\,i \\
Polar & 0.3736(36) - 0.0889(40)\,i & 0.599443 - 0.092703\,i & 0.809178 - 0.094164\,i \\
\end{tabular}
\caption{QNM frequencies (WKB method) in the commutative limit \(a\,m=0\), for \(n=0\) and \(M=1\,(R=2)\).}
\label{tab0a}
\end{table}
\par
When a small degree of noncommutativity is introduced, say \(a m = -0.001\), one begins to see a mismatch between the axial and polar spectra at higher \(\ell\). Table~\ref{tab0b} illustrates this effect:
% \begin{table}[h!]
% \begin{tabular}{m{4em} m{10em} m{10em} m {10em} }
% %\hline
% \\
%  & $\ell=2$ & $\ell=3$ &  $\ell=4$ \\
% %\\
% \hline
% \\
% Axial & 0.3736(38) - 0.0888(91) i & 0.599456 - 0.092703 i & 0.809188 - 0.094164 i \\
% Polar & 0.3736(58) - 0.0889(67) i & 0.599456 - 0.092716 i & 0.809187 - 0.094171 i \\
% \\
% \hline
% \end{tabular}
% \caption{QNM frequencies (WKB method) in the NC case \(a m=-0.001\), for \(n=0\) and \(M=1\,(R=2)\).}
% \label{tab0b}
% \end{table}
% %
% \par
\begin{table}[h!]
\begin{tabular}{m{4em} m{10em} m{10em} m{10em}}
 & $\ell=2$ & $\ell=3$ & $\ell=4$ \\
\hline
Axial & 0.3736(38) - 0.0888(91)\,i & 0.599456 - 0.092703\,i & 0.809188 - 0.094164\,i \\
Polar & 0.3736(58) - 0.0889(67)\,i & 0.599456 - 0.092716\,i & 0.809187 - 0.094171\,i \\
\end{tabular}
\caption{QNM frequencies (WKB method) in the NC case \(a\,m=-0.001\), for \(n=0\) and \(M=1\,(R=2)\).}
\label{tab0b}
\end{table}
\par
As one further increases the noncommutativity parameter, for example to \(a m = -0.01\), the loss of isospectrality grows more pronounced and is already visible for the fundamental mode at \(\ell=2\); see Table~\ref{tab0c}:
%
% \begin{table}[h!]
% \begin{tabular}{m{4em} m{10em} m{10em} m {10em} }
% %\hline
% \\
%  & $\ell=2$ & $\ell=3$ &  $\ell=4$ \\
% %\\
% \hline
% \\
% Axial & 0.3738(07) - 0.0888(92) i & 0.599573 - 0.092706 i & 0.809279 - 0.094166 i \\
% Polar & 0.3738(64) - 0.0892(07) i & 0.599571 - 0.092828 i & 0.809269 - 0.094238 i \\
% \\
% \hline
% \end{tabular}
% \caption{QNM frequencies from the WKB method in the NC case \(a m=-0.01\), for \(n=0\) and \(M=1\,(R=2).\)}
% \label{tab0c}
% \end{table}
% %
% \par
\begin{table}[h!]
\begin{tabular}{m{4em} m{10em} m{10em} m{10em}}
 & $\ell=2$ & $\ell=3$ & $\ell=4$ \\
\hline
Axial & 0.3738(07) - 0.0888(92)\,i & 0.599573 - 0.092706\,i & 0.809279 - 0.094166\,i \\
Polar & 0.3738(64) - 0.0892(07)\,i & 0.599571 - 0.092828\,i & 0.809269 - 0.094238\,i \\
\end{tabular}
\caption{QNM frequencies from the WKB method in the NC case \(a\,m=-0.01\), for \(n=0\) and \(M=1\,(R=2).\)}
\label{tab0c}
\end{table}
\par
Such isospectrality breaking persists across all negative values of \(a m\), and it also appears for positive \(a m\). To quantify the degree of this breaking, we define \cite{Liu:2024oeq}
\begin{equation}
    \Delta \omega _{R,I}=100 \times \frac{\omega _{R,I}^{\text{axial}}-\omega _{R,I}^{\text{polar}}}{\omega _{R,I}^{\text{polar}}},
\end{equation}
where the real and imaginary parts of the frequency are labeled \(\omega_R\) and \(\omega_I\). In Fig.~\ref{figiso6}, we show the results for this measure:
\begin{figure}[h!]
\centering
\subfigure[ref2][]{\includegraphics[scale=0.75]{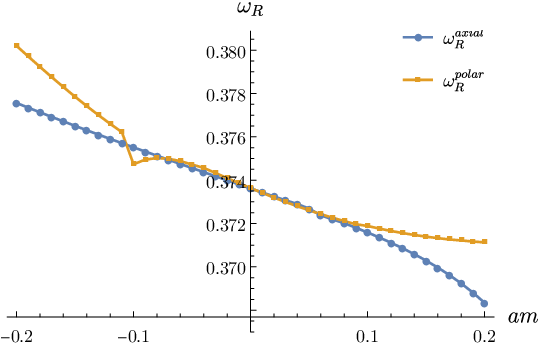}}
\qquad
\subfigure[ref3][]{\includegraphics[scale=0.75]{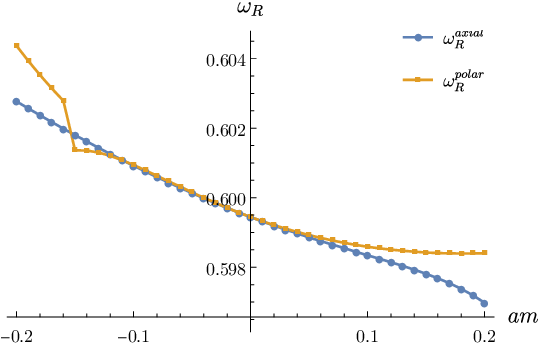}}
\subfigure[ref3][]{\includegraphics[scale=0.75]{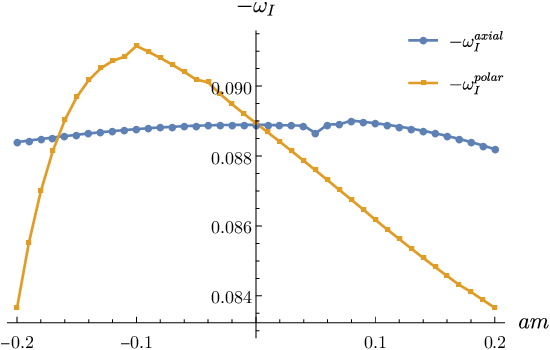}}
\qquad
\subfigure[ref3][]{\includegraphics[scale=0.75]{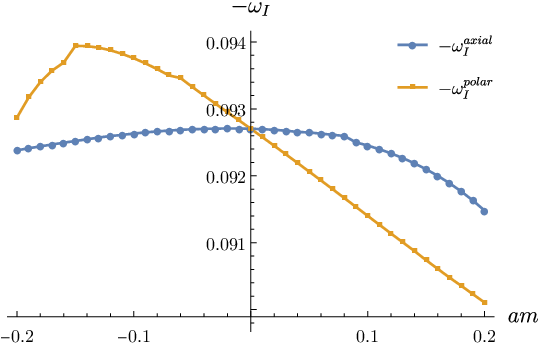}}
\subfigure[ref2][]{\includegraphics[scale=0.75]{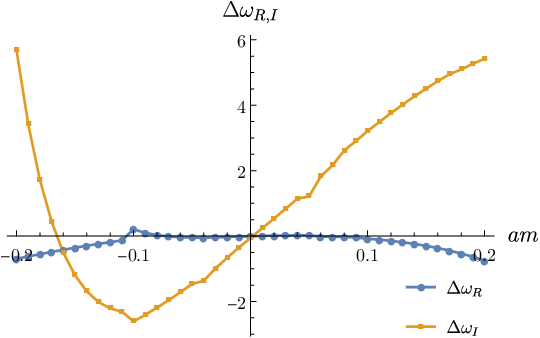}}
\qquad
\subfigure[ref3][]{\includegraphics[scale=0.75]{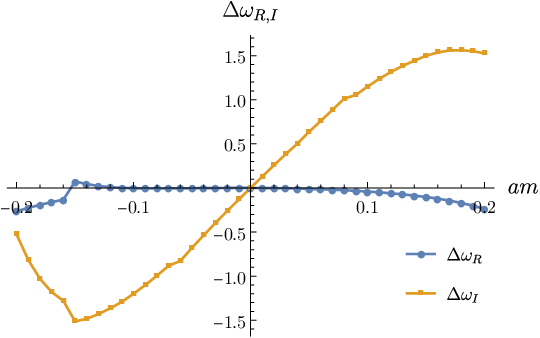}}
\caption{Illustration of the isospectrality breaking induced by noncommutativity. \emph{Left column:} \(\ell=2\). \emph{Right column:} \(\ell=3\). The first row plots the real part \(\omega_R\) versus \(a\,m\). The second row shows the imaginary part \(\omega_I\). The bottom row displays the relative deviation \(\Delta \omega_{R,I}\). In each panel, we fix \(n=0\), \(M=1\,(R=2)\), and choose the QNM values at the optimal WKB order.}
\label{figiso6}
\end{figure}
Firstly, one observes that the real part of the quasinormal frequency is affected in a similar way for both axial and polar modes, with notable deviations only setting in once \(\abs{a m}\) becomes large enough to challenge the WKB precision. Secondly, the imaginary part \(\omega_I\) behaves differently: for a given \(a m\), it undergoes changes that run counter to those of the real part, and in the axial case, \(\omega_I\) remains comparatively less sensitive to noncommutativity. These patterns indicate that the NC deformation modifies both the oscillation frequencies and damping times, though not in the same degree for the two parities. Indeed, the isospectrality violation tends to be more evident in the imaginary component.

A qualitative reason behind the stronger influence on polar modes can be traced to the nature of the twist in \eqref{kxdef}, specifically the quantization of the radial coordinate \(r\). Axial perturbations encode off-diagonal components of the metric perturbation, usually associated with the “twisting” or shear-like deformations of the spacetime, akin to gravitational waves generated by orbiting or spiraling matter. By contrast, polar perturbations modify the radial and time–time components, effectively capturing spherically symmetric changes in distance measures—the so-called “breathing-like” oscillations of the black hole. Because the radial coordinate is directly quantized in this NC setup, polar modes are more strongly impacted by the resulting quantum-geometric corrections.

\section{Conclusion}
We have investigated the influence of a noncommutative (NC) deformation on classical Schwarzschild black hole perturbations, unveiling new insights into how quantum-gravity-inspired effects may manifest in a ringdown phase. By employing a Drinfeld twist in the differential geometry of spacetime, we derived NC analogs of the Regge-Wheeler and Zerilli equations, finding that their usual isospectrality is no longer upheld. Such a spectral splitting between axial and polar QNMs serves as a distinct theoretical signature of noncommutativity.

To probe these effects numerically, we conducted semi-analytical calculations using higher-order WKB approximations, as well as Pöschl–Teller and Rosen–Morse fitting techniques. Our analysis shows that NC corrections can be consistently incorporated into black hole perturbation theory, permitting reliable estimates for the new quasinormal frequencies. While these frequency shifts are generally small, advanced gravitational-wave detectors of the future might discern such tiny deviations, offering a path to testing quantum-gravitational scenarios. Moreover, our results demonstrate that the formalism is readily extensible, suggesting broader applications to rotating black holes, higher-dimensional models, and other effective theories beyond General Relativity. Ultimately, these findings underscore the value of searching for subtle signatures of NC geometry as part of the continuing quest to uncover viable routes toward quantum gravity.\\

\noindent{\bf Acknowledgment}\\
This  research was supported by the Croatian Science Foundation Project No. IP-2020-02-9614 \textit{Search for Quantum spacetime in Black Hole QNM spectrum and Gamma Ray Bursts}. Part of the calculations were checked using the Mathematica package.

\bibliography{BibTex}

\end{document}